\documentclass[%
reprint,
nofootinbib,
amsmath,
amssymb,
aps, 
physrev,
]{revtex4-2}
\usepackage{bm}
\usepackage{booktabs}
\usepackage{braket}
\usepackage{dcolumn}
\usepackage{graphicx}
\usepackage{hyperref}
\usepackage{mhchem}
\usepackage{siunitx}
\newcommand*\dif{\mathop{}\!\mathrm{d}}

\begin{document}
\title{A Generalized Multinodal Model for Plasma Particle and Energy Transport}
\author{Zefang Liu}\email{Contact author: liuzefang@gatech.edu}
\author{Weston M. Stacey}
\affiliation{
Fusion Research Center, Georgia Institute of Technology, Atlanta, GA, USA
}
\begin{abstract}
We present a generalized multinodal model for simulating particle and energy transport in toroidal plasma configurations, developed to support burning plasma analysis and reactor-scale modeling. Unlike fixed-node models, this formulation allows an arbitrary number of nodes, offering increased flexibility for coupling with core-edge or core-pedestal simulations. The model derives nodal balance equations for each plasma species by volume-averaging the continuity and energy conservation equations across toroidal shell nodes. Particle and energy transport terms are expressed in terms of internodal fluxes, linked to radial gradients via linear diffusion laws for particle density and temperature, respectively. The resulting transport contributions are characterized through effective particle and energy transport times, derived explicitly in terms of nodal geometry and diffusivities. This generalized framework facilitates efficient, modular implementation of radial transport dynamics in reduced-order or integrated plasma simulations, and is compatible with data-driven approaches such as NeuralPlasmaODE for model calibration and inference from experimental data.
\end{abstract}
\maketitle
\section{Introduction}

The realization of sustained fusion energy requires accurate modeling of burning plasma behavior in reactor-scale devices such as ITER~\cite{aymar2002iter,green2003iter,holtkamp2007overview}. Burning plasmas, in which the majority of heating arises from fusion-born alpha particles, involve tightly coupled dynamics across space and time scales, necessitating tractable models that retain essential physical fidelity~\cite{wang1997simulation,cordey2005scaling}. While high-fidelity simulations offer detailed insight, reduced-order nodal models have emerged as valuable tools for studying global plasma dynamics and facilitating real-time control development~\cite{stacey2021nodal,hill2017confinement,hill2019burn}. These models discretize the plasma volume into a set of nodes, typically concentric shells, allowing volume-averaged equations to capture key transport phenomena.

Building upon previous work on multi-region multi-timescale nodal frameworks~\cite{stacey2021nodal,liu2020one,liu2021multi,liu2022multi,liu2022thesis}, we present a generalized multinodal model in which the number and structure of nodes are arbitrary. This flexibility enables modular integration with core, edge, or scrape-off layer (SOL) models and supports adaptive spatial resolution based on local physics needs. The formulation derives volume-averaged balance equations for particles and energy by applying fluid theory in toroidal geometry~\cite{stacey2012fusion,wesson2011tokamaks}. Transport is expressed in terms of internodal fluxes, linked to gradients via linear diffusion laws, and cast into effective transport times that scale naturally with node geometry. Such an approach provides a systematic path toward scalable and interpretable reduced-order models, complementary to modern machine learning-based dynamical models such as NeuralPlasmaODE\footnote{\url{https://github.com/zefang-liu/NeuralPlasmaODE}}~\cite{liu2024application,liu2024application2,liu2025sensitivity,liu2025optimizing} based on Neural Ordinary Differential Equations (Neural ODEs)~\cite{chen2018neural}.

This general formulation is intended to support ongoing efforts to understand and control complex burning plasma phenomena, including fueling, impurity transport, and thermal equilibration. By maintaining physical consistency while allowing structural flexibility, the model can be readily embedded into integrated simulations or control-oriented frameworks targeting reactor operation and optimization.

\section{Model Geometry}

\begin{figure}[!h]
\centering
\includegraphics[width=\linewidth]{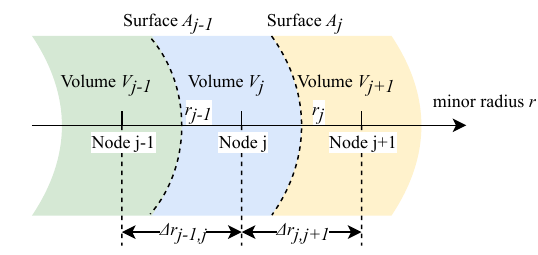}
\caption{Schematic of the multinodal model, illustrating toroidal shell nodes and internodal torus surfaces.}
\label{fig:nodes}
\end{figure}

The geometry of the generalized multinodal model is illustrated in Figure~\ref{fig:nodes}. In this framework, the plasma domain is discretized into a series of concentric toroidal shells (nodes), separated by toroidal surfaces (internodal interfaces). Let $r_j$ denote the minor radius of the surface $A_j$, and $\Delta r_{j,j+1}$ the radial distance between adjacent nodes $j$ and $j+1$.

We adopt standard notation for plasma species: the set of ion species is given by $\mathcal{I} = \set{\ce{D}, \ce{T}, \alpha, z_1, z_2, \dots}$, and the full set of species is $\mathcal{S} = \set{e} \cup \mathcal{I}$. The formulation begins with the derivation of particle balance equations, followed by the corresponding energy balance equations for each species.

\section{Balance Equations}

To model the spatiotemporal evolution of plasma quantities within the multinodal framework, we derive volume-averaged balance equations for each species. These include particle and energy conservation equations, formulated to capture internodal transport and source effects in toroidal geometry.

\subsection{Particle Balance Equations}

The particle transport for each species $\sigma \in \mathcal{S}$ is governed by the continuity equation from fluid theory~\cite{stacey2012fusion}:
\begin{equation}
    \frac{\partial n_\sigma}{\partial t} + \nabla \cdot \bm{\Gamma}_\sigma = S_{\sigma},
\end{equation}
where $n_\sigma$ is the particle density, $\bm{\Gamma}_\sigma = n_\sigma \bm{v}_\sigma$ is the particle flux, and $S_\sigma$ is the net particle source.

\subsubsection{Volume-Averaged Particle Balance}

Averaging over the volume $V_j$ of node $j$, we obtain:
\begin{equation}
    \frac{1}{V_j} \int_{V_j} \frac{\partial n_\sigma}{\partial t} \dif V = \frac{1}{V_j} \int_{V_j} S_{\sigma} \dif V - \frac{1}{V_j} \int_{V_j} \nabla \cdot \bm{\Gamma}_\sigma \dif V.
\end{equation}
We define the volume-averaged quantities:
\begin{align}
    \bar{n}_{\sigma}^{(j)} &= \frac{1}{V_j} \int_{V_j} n_{\sigma} \, \dif V, \\
    \bar{S}_{\sigma}^{(j)} &= \frac{1}{V_j} \int_{V_j} S_{\sigma} \, \dif V, \\
    \bar{S}_{\sigma, \text{tran}}^{(j)} &= - \frac{1}{V_j} \int_{V_j} \nabla \cdot \bm{\Gamma}_\sigma \, \dif V.
\end{align}
This yields the nodal (volume-averaged) particle balance equation:
\begin{equation}
    \frac{\dif \bar{n}_\sigma^{(j)}}{\dif t} = \bar{S}_{\sigma}^{(j)} + \bar{S}_{\sigma, \text{tran}}^{(j)}.
\end{equation}

\subsubsection{Particle Flux Approximation}

Applying the divergence theorem, the transport term becomes:
\begin{equation}
    \bar{S}_{\sigma, \text{tran}}^{(j)} = - \frac{1}{V_j} \left( \int_{A_j} \bm{\Gamma}_\sigma \cdot \dif \bm{S} - \int_{A_{j-1}} \bm{\Gamma}_\sigma \cdot \dif \bm{S} \right).
\end{equation}
Assuming Fick's law~\cite{vincenti1965introduction} $\bm{\Gamma}_\sigma = - D_\sigma \nabla n_\sigma$, the density gradient in toroidal coordinates~\cite{stacey2012fusion} is:
\begin{equation}
    \nabla n_\sigma = \frac{\partial n_\sigma}{\partial r} \bm{\hat{r}} + \frac{1}{R_0 + r \cos \theta} \frac{\partial n_\sigma}{\partial \phi} \bm{\hat{\phi}} + \frac{1}{r} \frac{\partial n_\sigma}{\partial \theta} \bm{\hat{\theta}}.
\end{equation}
Assuming toroidal and poloidal symmetry of $n_\sigma$ across each internodal surface, we retain only the radial component:
\begin{equation}
    \nabla n_\sigma \approx \frac{\dif n_\sigma}{\dif r} \bm{\hat{r}}.
\end{equation}
Then, the transport term simplifies to:
\begin{equation}
    \bar{S}_{\sigma, \text{tran}}^{(j)} = - \frac{1}{V_j} \left[ \left( \Gamma_{\sigma, r} \right)_{A_j} A_j - \left( \Gamma_{\sigma, r} \right)_{A_{j-1}} A_{j-1} \right],
\end{equation}
where the radial fluxes are approximated using finite differences:
\begin{align}
    \left( \Gamma_{\sigma, r} \right)_{A_j} &\approx - D_{\sigma}^{(j)} \frac{\bar{n}_{\sigma}^{(j+1)} - \bar{n}_{\sigma}^{(j)}}{\Delta r_{j,j+1}}, \\
    \left( \Gamma_{\sigma, r} \right)_{A_{j-1}} &\approx - D_{\sigma}^{(j-1)} \frac{\bar{n}_{\sigma}^{(j)} - \bar{n}_{\sigma}^{(j-1)}}{\Delta r_{j-1,j}}.
\end{align}

\subsubsection{Internodal Particle Transport Times}

To express transport in a time-scale form, we define the internodal particle transport times:
\begin{align}
    \tau_{P, \sigma}^{j \to j+1} &= \frac{V_j \Delta r_{j,j+1}}{A_j D_{\sigma}^{(j)}} = \frac{r_j^2 - r_{j-1}^2}{2r_j} \frac{\Delta r_{j,j+1}}{D_{\sigma}^{(j)}}, \\
    \tau_{P, \sigma}^{j+1 \to j} &= \frac{V_{j+1} \Delta r_{j,j+1}}{A_j D_{\sigma}^{(j)}} = \frac{r_{j+1}^2 - r_j^2}{2r_j} \frac{\Delta r_{j,j+1}}{D_{\sigma}^{(j)}}, \\
    \tau_{P, \sigma}^{j-1 \to j} &= \frac{V_{j-1} \Delta r_{j-1,j}}{A_{j-1} D_{\sigma}^{(j-1)}} = \frac{r_{j-1}^2 - r_{j-2}^2}{2r_{j-1}} \frac{\Delta r_{j-1,j}}{D_{\sigma}^{(j-1)}}, \\
    \tau_{P, \sigma}^{j \to j-1} &= \frac{V_j \Delta r_{j-1,j}}{A_{j-1} D_{\sigma}^{(j-1)}} = \frac{r_j^2 - r_{j-1}^2}{2r_{j-1}} \frac{\Delta r_{j-1,j}}{D_{\sigma}^{(j-1)}}.
\end{align}

Here, $V_j = 2\pi R_0 \cdot \pi (r_j^2 - r_{j-1}^2)$ is the toroidal shell volume, and $A_j = 2\pi R_0 \cdot 2\pi r_j$ is the toroidal surface area. For elongated plasmas, $r_j^2$ can be replaced by $a_j b_j$, where $a_j$ and $b_j$ are the semi-major and semi-minor radii of the poloidal cross-section.

The transport term can now be written as:
\begin{equation}
    \bar{S}_{\sigma, \text{tran}}^{(j)} = - \frac{\bar{n}_{\sigma}^{(j)} - \bar{n}_{\sigma}^{(j+1)}}{\tau_{P, \sigma}^{j \to j+1}} - \frac{\bar{n}_{\sigma}^{(j)} - \bar{n}_{\sigma}^{(j-1)}}{\tau_{P, \sigma}^{j \to j-1}}.
\end{equation}
Alternatively, using the relationships
\begin{align}
    \tau_{P, \sigma}^{j+1 \to j} &= \frac{V_{j+1}}{V_j} \tau_{P, \sigma}^{j \to j+1}, \\
    \tau_{P, \sigma}^{j-1 \to j} &= \frac{V_{j-1}}{V_j} \tau_{P, \sigma}^{j \to j-1},
\end{align}
we may express the transport term in symmetric form:
\begin{equation}
\begin{split}
    \bar{S}_{\sigma, \text{tran}}^{(j)} 
    &= - \frac{\bar{n}_{\sigma}^{(j)}}{\tau_{P, \sigma}^{j \to j+1}} 
    + \frac{V_{j+1}}{V_j} \frac{\bar{n}_{\sigma}^{(j+1)}}{\tau_{P, \sigma}^{j+1 \to j}} 
    \\
    &\quad - \frac{\bar{n}_{\sigma}^{(j)}}{\tau_{P, \sigma}^{j \to j-1}} 
    + \frac{V_{j-1}}{V_j} \frac{\bar{n}_{\sigma}^{(j-1)}}{\tau_{P, \sigma}^{j-1 \to j}}.
\end{split}
\end{equation}

\subsection{Energy Balance Equations}

The conservation of energy for each species $\sigma \in \mathcal{S}$ is governed by the fluid energy equation~\cite{freidberg2008plasma}, which accounts for compressional heating, convective transport, and thermal conduction:
\begin{equation}
    \frac{3}{2} n_\sigma \left( \frac{\partial}{\partial t} + \bm{v}_\sigma \cdot \nabla \right) T_\sigma + p_\sigma \nabla \cdot \bm{v}_\sigma + \nabla \cdot \bm{q}_\sigma = P_\sigma,
\end{equation}
where $T_\sigma$ is the temperature (in energy units), $p_\sigma$ the pressure, $\bm{q}_\sigma$ the heat flux, and $P_\sigma$ the net energy source for species $\sigma$.

\subsubsection{Energy Density Reformulation}

Defining the energy density as $U_\sigma = \frac{3}{2} n_\sigma T_\sigma$, and combining the energy and particle balance equations, the conservation law becomes:
\begin{align}
\begin{split}
    \frac{3}{2} n_\sigma \frac{\partial T_\sigma}{\partial t}
    &+ \frac{3}{2} \bm{\Gamma}_\sigma \cdot \nabla T_\sigma
    + p_\sigma \nabla \cdot \bm{v}_\sigma
    + \nabla \cdot \bm{q}_\sigma \\
    &+ \frac{3}{2} T_\sigma \frac{\partial n_\sigma}{\partial t}
    + \frac{3}{2} T_\sigma \nabla \cdot \bm{\Gamma}_\sigma
    = P_\sigma,
\end{split} \\
\begin{split}
    \frac{3}{2} \frac{\partial (n_\sigma T_\sigma)}{\partial t}
    & = P_\sigma - p_\sigma \nabla \cdot \bm{v}_\sigma \\
    &\quad - \nabla \cdot \left( \frac{3}{2} \bm{\Gamma}_\sigma T_\sigma + \bm{q}_\sigma \right).
\end{split}
\end{align}
The particle and heat fluxes are modeled as
\begin{align}
    \bm{\Gamma}_\sigma &= - D_\sigma \nabla n_\sigma, \\
\begin{split}
    \bm{q}_\sigma &= - k_\sigma \nabla \left( \frac{T_\sigma}{k} \right) = - \chi_\sigma n_\sigma c_{p,m,\sigma} \nabla \left( \frac{T_\sigma}{k} \right) \\
    &= - \frac{5}{2} \chi_\sigma n_\sigma \nabla T_\sigma,
\end{split}
\end{align}
where $D_\sigma$ is the diffusion coefficient, $k_\sigma$ the thermal conductivity, $\chi_\sigma$ the thermal diffusivity, and $c_{p,m,\sigma}$ the molar heat capacity. The heat flux expression is based on Fourier's law~\cite{cengel2019thermodynamics} of thermal conduction. Substituting these forms, the energy equation becomes:
\begin{equation}
\begin{split}
    \frac{\partial U_\sigma}{\partial t} &= \left( P_\sigma - p_\sigma \nabla \cdot \bm{v}_\sigma \right) \\
    &\quad + \nabla \cdot \left( \frac{3}{2} D_\sigma T_\sigma \nabla n_\sigma 
    + \frac{5}{2} \chi_\sigma n_\sigma \nabla T_\sigma \right).
\end{split}
\end{equation}

Assuming $D_\sigma \approx \frac{5}{3} \chi_\sigma$~\cite{becker2006anomalous}, and incorporating the compressional term into the effective source, we redefine:
\begin{equation}
    \chi_\sigma \leftarrow \frac{5}{3} \chi_\sigma \approx D_\sigma, \quad 
    P_\sigma \leftarrow P_\sigma - p_\sigma \nabla \cdot \bm{v}_\sigma.
\end{equation}
This simplifies the energy conservation law to:
\begin{equation}
\begin{split}
    \frac{\partial U_\sigma}{\partial t} 
    &= P_\sigma + \nabla \cdot \left( \frac{3}{2} \chi_\sigma T_\sigma \nabla n_\sigma 
    + \frac{3}{2} \chi_\sigma n_\sigma \nabla T_\sigma \right) \\
    &= P_\sigma + \nabla \cdot \left( \chi_\sigma \nabla U_\sigma \right).
\end{split}
\end{equation}

\subsubsection{Volume-Averaged Energy Balance}

Following the same nodal approach as for particles, we define the volume-averaged quantities:
\begin{align}
    \bar{U}_\sigma^{(j)} &= \frac{1}{V_j} \int_{V_j} U_\sigma \, \dif V, \\
    \bar{P}_\sigma^{(j)} &= \frac{1}{V_j} \int_{V_j} P_\sigma \, \dif V, \\
    \bar{P}_{\sigma, \text{tran}}^{(j)} &= \frac{1}{V_j} \int_{V_j} \nabla \cdot \left( \chi_\sigma \nabla U_\sigma \right) \, \dif V.
\end{align}
The nodal energy balance equation becomes:
\begin{equation}
    \frac{\dif \bar{U}_\sigma^{(j)}}{\dif t} = \bar{P}_\sigma^{(j)} + \bar{P}_{\sigma, \text{tran}}^{(j)},
\end{equation}
with the energy density expressed as:
\begin{equation}
    \bar{U}_\sigma^{(j)} = \frac{3}{2} \bar{n}_\sigma^{(j)} \bar{T}_\sigma^{(j)}.
\end{equation}

\subsubsection{Internodal Energy Transport Times}

Using the same geometric framework as in the particle model, we define the energy transport times:
\begin{align}
    \tau_{E, \sigma}^{j \to j+1} &= \frac{V_j \Delta r_{j,j+1}}{A_j \chi_\sigma^{(j)}} = \frac{r_j^2 - r_{j-1}^2}{2r_j} \frac{\Delta r_{j,j+1}}{\chi_\sigma^{(j)}}, \\
    \tau_{E, \sigma}^{j+1 \to j} &= \frac{V_{j+1} \Delta r_{j,j+1}}{A_j \chi_\sigma^{(j)}} = \frac{r_{j+1}^2 - r_j^2}{2r_j} \frac{\Delta r_{j,j+1}}{\chi_\sigma^{(j)}}, \\
    \tau_{E, \sigma}^{j-1 \to j} &= \frac{V_{j-1} \Delta r_{j-1,j}}{A_{j-1} \chi_\sigma^{(j-1)}} = \frac{r_{j-1}^2 - r_{j-2}^2}{2r_{j-1}} \frac{\Delta r_{j-1,j}}{\chi_\sigma^{(j-1)}}, \\
    \tau_{E, \sigma}^{j \to j-1} &= \frac{V_j \Delta r_{j-1,j}}{A_{j-1} \chi_\sigma^{(j-1)}} = \frac{r_j^2 - r_{j-1}^2}{2r_{j-1}} \frac{\Delta r_{j-1,j}}{\chi_\sigma^{(j-1)}}.
\end{align}

Finally, the volume-averaged energy transport term is written as:
\begin{equation}
\begin{split}
    \bar{P}_{\sigma, \text{tran}}^{(j)} 
    = - \frac{\bar{U}_\sigma^{(j)} - \bar{U}_\sigma^{(j+1)}}{\tau_{E, \sigma}^{j \to j+1}} 
    - \frac{\bar{U}_\sigma^{(j)} - \bar{U}_\sigma^{(j-1)}}{\tau_{E, \sigma}^{j \to j-1}}.
\end{split}
\end{equation}
Alternatively, this can be expressed symmetrically as:
\begin{equation}
\begin{split}
    \bar{P}_{\sigma, \text{tran}}^{(j)} 
    &= - \frac{\bar{U}_\sigma^{(j)}}{\tau_{E, \sigma}^{j \to j+1}} 
    + \frac{V_{j+1}}{V_j} \frac{\bar{U}_\sigma^{(j+1)}}{\tau_{E, \sigma}^{j+1 \to j}} \\
    &\quad - \frac{\bar{U}_\sigma^{(j)}}{\tau_{E, \sigma}^{j \to j-1}} 
    + \frac{V_{j-1}}{V_j} \frac{\bar{U}_\sigma^{(j-1)}}{\tau_{E, \sigma}^{j-1 \to j}}.
\end{split}
\end{equation}

\section{Conclusion}

We presented a generalized multinodal framework for modeling particle and energy transport in toroidal plasma systems. The model is formulated using volume-averaged balance equations for multiple species, incorporating diffusive transport and representing internodal coupling through effective transport times. Its flexible structure supports arbitrary nodal discretizations and species sets, making it broadly applicable to diverse plasma configurations and reduced-order modeling tasks.

While this work focuses on the theoretical formulation, the model is designed to integrate naturally with data-driven frameworks such as NeuralPlasmaODE~\cite{liu2024application,liu2024application2,liu2025sensitivity,liu2025optimizing}. In that context, transport times and diffusivities~\cite{becker2004study,becker2006anomalous} can be treated as learnable parameters and inferred from experimental profile data. This capability enables data-informed calibration while maintaining physical consistency, providing a foundation for scalable, interpretable, and predictive modeling of burning plasma dynamics.

\bibliography{apssamp}
\end{document}